\documentclass[10pt,letterpaper]{article}
\usepackage{opex3}
\usepackage{graphicx}
\DeclareGraphicsExtensions{.pdf,.jpeg,.png,.eps}
\usepackage{epstopdf}
\usepackage[font=footnotesize]{subfig}
\usepackage[cmex10]{amsmath}
\usepackage{xcolor}
\usepackage{siunitx}
\usepackage{cite}
\usepackage{mathptmx}
\usepackage{courier}
\usepackage{helvet}
\usepackage{soul}
\usepackage[none]{hyphenat}

\begin{document}

\title{Wavelength tuning and stabilization of microring-based filters using silicon in-resonator photoconductive heaters }

\author{Hasitha Jayatilleka,* Kyle Murray, Miguel \'{A}ngel Guill\'{e}n-Torres,  \\ Michael Caverley, Ricky Hu, Nicolas A. F. Jaeger,  Lukas Chrostowski, \\ and Sudip Shekhar}

\address{Department of Electrical and Computer Engineering, University of British Columbia. \\
2332 Main Mall, Vancouver, British Columbia V6T 1Z4, Canada.}

\email{*hasitha@ece.ubc.ca} 



\begin{abstract}
We demonstrate that $n$-doped resistive heaters in silicon waveguides show photoconductive effects with high responsivities. 
These photoconductive heaters, integrated into microring resonator (MRR)-based filters, were used to automatically tune and stabilize the filter's resonance wavelength to the input laser's wavelength.
This is achieved without requiring dedicated defect implantations, additional material depositions, dedicated photodetectors, or optical power tap-outs. 
Automatic wavelength stabilization of first-order MRR and second-order series-coupled MRR filters is experimentally demonstrated. Open eye diagrams were obtained for data transmission at \SI{12.5}{Gb/s} while the temperature was varied by 5 $^\circ$C at a rate of 0.28 $^\circ$C/s.
We theoretically show that series-coupled MRR-based  filters of any order can be automatically tuned by using photoconductive heaters to monitor the light intensity in each MRR, and sequentially aligning the resonance of each MRR to the laser's wavelength. 
\end{abstract}



\section{Introduction}
Silicon photonic microring resonator (MRR)-based filters, modulators, and switches have been investigated for use in data-centers and high performance computing systems due to their high-speed operation, low power consumption, and compact device footprints \cite{ophir2013}.
Many of the benefits exhibited by MRR-based devices are due to their resonance characteristics,
which also makes their performance highly susceptible to variations in fabrication and to fluctuations in both the chip temperature and  the laser's wavelength. 
In order to overcome these issues for practical deployment of MRR-based technologies, it is required to develop scalable, low-cost, and energy efficient techniques for wavelength tuning and stabilization of MRR-based devices \cite{padmaraju2014_review, morichetti2014}.

Automatic wavelength tuning and stabilization of MRR-based devices is typically achieved using feedback loops, which require both sensing and controlling the resonance conditions of the MRRs.
The sensing operation can be performed using on-chip temperature sensors \cite{derose2010} or, alternatively, using photodetectors (PDs) to monitor the light intensity at the output ports of \cite{cox2014, mak2015}, or inside \cite{zhang2014, li2015, poulton2015, grillanda2014}, the MRRs. 
Techniques that require light to be tapped out from the MRRs or their outputs in order to be monitored with on-chip PDs do not scale well towards densely integrated systems due to the increases in  device footprint and insertion losses. 
In contrast, sensing mechanisms based on monitoring the light intensity in the MRRs with in-resonator PDs are more scalable. For example, in-resonator PDs can be fabricated that utilize defect state absorption (DSA) as the photodetection mechanism \cite{zhang2014, li2015, fard2014}. DSA is the process of electron-hole pair generation by sub-bandgap defect energy levels, formed primarily as a result of ion implantation \cite{yu2012, zhang2014, logan2011}. Previous demonstrations have required dedicated ion implantation steps to create a sufficient number of defect states for absorption \cite{zhang2014, li2015}.

The control operation can be performed using thermo-optic (TO) \cite{padmaraju2014}, or electro-optic~(EO) \cite{chengli2014}, phase shifters to tune the resonance wavelength of the MRRs. 
The TO phase shifters are typically implemented using  metallic \cite{padmaraju2014, grillanda2014} or doped-silicon \cite{timurdogan2012, li2015, de_heyn2013} resistive heating elements and are widely used for tuning MRRs due to the large TO coefficient of silicon. 
For wavelength stabilization, these heating elements are typically used together with dedicated PDs inside or outside of the MRRs. 
Recently, germanium-based in-resonator photoconductive heaters have been demonstrated for both sense and control operations \cite{poulton2015}, thereby avoiding the need for dedicated PDs. 
However, these devices required the deposition of germanium, increasing the complexity of fabrication. 
The TO or EO tuners can be  driven by microcontrollers \cite{timurdogan2012, padmaraju2014, li2015} or CMOS electronic circuits \cite{grillanda2014, chengli2014, zheng2014}.
The control algorithms for wavelength stabilization can be based on thermal dithering \cite{padmaraju2014}, homodyne locking \cite{cox2014}, or maximum/minimum point searching \cite{timurdogan2012, zheng2014} techniques.

In this paper, we demonstrate automated wavelength tuning and stabilization of MRR filters using in-resonator photoconductive heaters (IRPHs). IRPHs are formed using doped waveguide sections [see Fig. \ref{fig1}(a)]. The doping allows the IRPHs to be used as resistive heaters, but also provide a means of photodetection due to DSA.  
In this work, the doped heaters built into the MRRs, i.e. IRPHs, are used for both the sense and the control operations. 
As a result, no dedicated ion implantation steps, germanium depositions, or dedicated on-chip PDs are required for automated stabilization of the MRRs.
Furthermore, the use of IRPHs allows for a simple and scalable method for tuning higher-order series-coupled MRR filters since the tuning state of an MRR can be measured using the MRR's IRPH.
All the devices described in our paper were fabricated using \SI{248}{nm} optical lithography at A*STAR IME.
To the best of our knowledge, this is the first demonstration of doped silicon IRPH-based automatic tuning and stabilization of first and second-order MRR-based filters.

The rest of this paper is organized as follows. In section 2, the photoconductive behaviors  of IRPHs are characterized.
In section 3, the wavelength stabilization of a first-order MRR filter using an IRPH is experimentally demonstrated. 
In section 4, automatic wavelength tuning and stabilization of a second-order series-coupled MRR filter is experimentally demonstrated.
In section 5, it is theoretically shown that IRPHs can be used to automatically tune higher-order series-coupled MRR-based filters to the input laser's wavelength.
Section 6 presents the discussion and the conclusions.

\section{In-resonator photoconductive heaters}

\begin{figure*}[h]
\centering
\centerline{
{\includegraphics[width=0.85\textwidth]{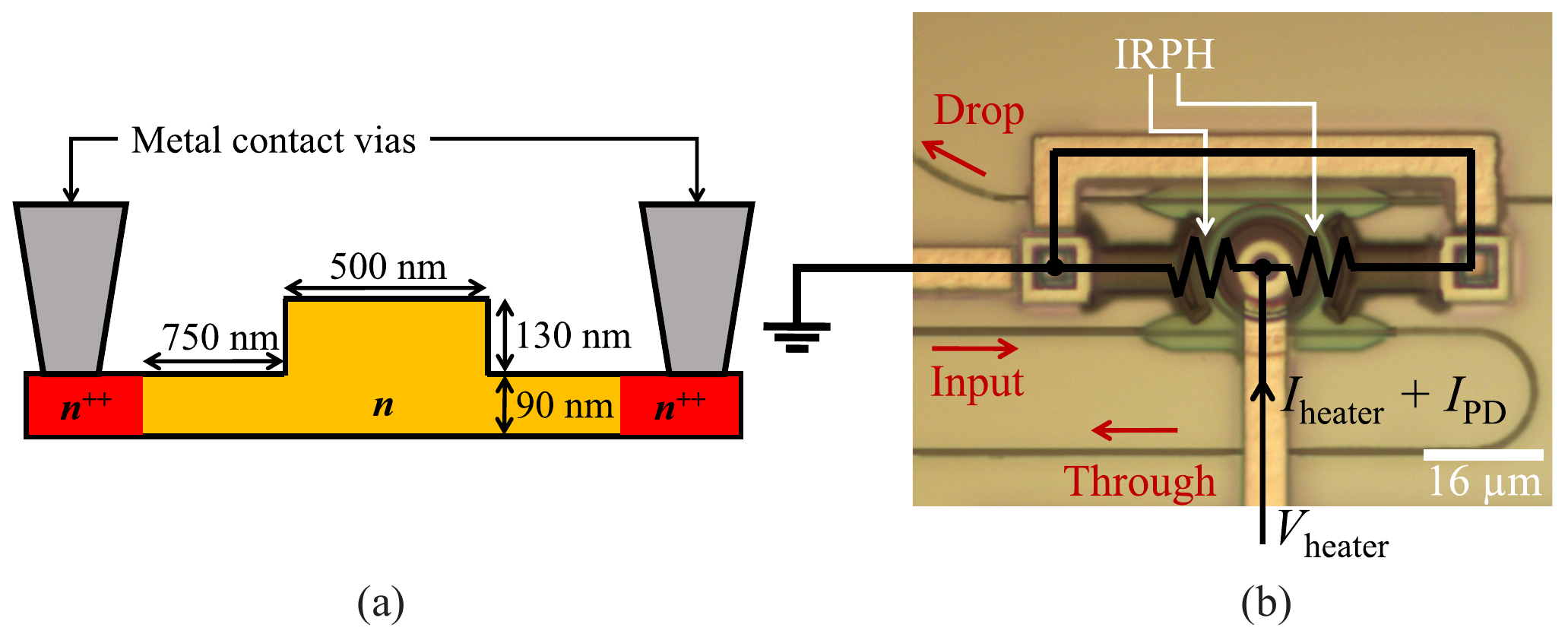}}
}
\caption{(a) Schematic cross-section of the waveguide used in the IRPH. (b) Microscope image of an add/drop MRR filter overlaid with a circuit description of the integrated IRPH.  
}
\label{fig1}
\end{figure*}

In this section, we characterize the photoconductive behavior of IRPHs, which can be used to monitor and control the light intensity in MRR-based filters. 
A schematic cross-section of the IRPH design used is shown in Fig. \ref{fig1}(a). The waveguide is n-doped and the $n^{++}$-doped regions on each side of the waveguide facilitate low resistance contact to the silicon.
Figure \ref{fig1}(b) shows a microscope image of a first-order add/drop MRR filter with integrated IRPHs. The radius of the MRR is \SI{8}{\um} and the IRPH is formed over 63\% of the MRR's circumference.
The MRR is symmetrically coupled, with identical power coupling coefficients of $|\kappa|^2 = 0.047$, corresponding to a gap of \SI{255}{nm}, for the through- and drop-port.
The total waveguide loss was approximately \SI{6.9}{dB/cm}. 
The values for $|\kappa|^2$ and waveguide loss were extracted from the measured MRR filter spectrum using a method similar to the one described in\cite{mckinnon2009}. 
The free-spectral-range of the fabricated device was \SI{12.4}{nm}.

\begin{figure*}[b]
\centering
\centerline{
{\includegraphics[height=1.7in]{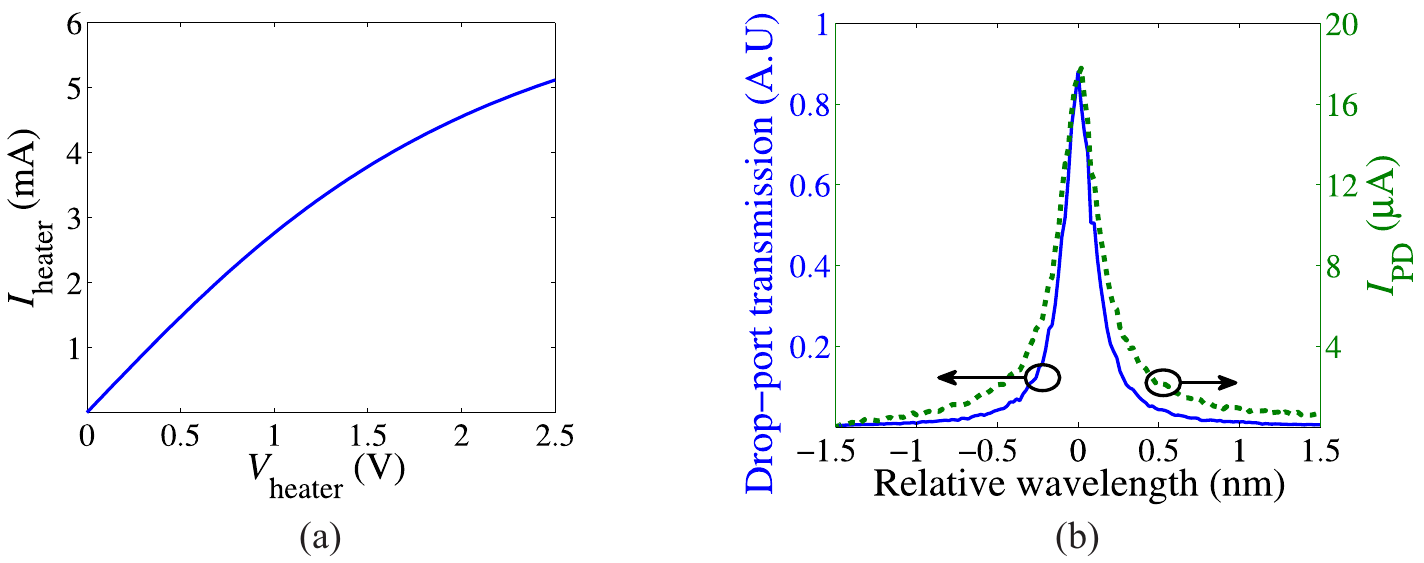}} \hspace{0.2cm}}
\caption{(a) Measured $I_{\text{heater}}$ as a function of $V_{\text{heater}}$. (b) Normalized drop-port transmission and $I_{\text{PD}}$ as a function of the wavelength offset relative to \SI{1551.52}{nm}. $V_{\text{heater}} = $ \SI{1}{V} and the drop-port transmission is normalized to \SI{87}{\uW}, which was the estimated optical power in the bus waveguide at the input of the MRR filter.  
}
\label{fig2}
\end{figure*}

The resonance wavelength of the MRR filter can be tuned by applying a voltage, $V_{\text{heater}}$, across the IRPH. The total current flowing consists of two parts: (1) the dark current, $I_{\text{heater}}$, which is the current that flows when no light is incident on the IRPH, and (2) the photocurrent, $I_{\text{PD}}$, which is the current generated due to DSA. $I_{\text{PD}}$ depends on both the light intensity inside the MRR and $V_{\text{heater}}$.
The measured $I_{\text{heater}}$ versus $V_{\text{heater}}$ for the MRR filter is shown in Fig.~\ref{fig2}(a). 
The electrical power supplied to the IRPH shifts the resonance wavelength of the MRR by \SI{0.25}{nm/mW}, or equivalently, changes the round-trip phase by \SI{0.04}{$\pi$/mW}. 
Figure \ref{fig2}(b) shows the measured drop-port optical transmission and $I_{\text{PD}}$ as a function of the wavelength of the optical input. 
For this measurement, $V_{\text{heater}} =$ \SI{1}{V}, and $I_{\text{PD}}$ was extracted by subtracting $I_{\text{heater}}$ from the total measured current. The same calibration method was used to obtain $I_{\text{PD}}$ values presented in the rest of this paper.

\begin{figure*}[h]
\centering
\centerline{
{\includegraphics[height=1.7in]{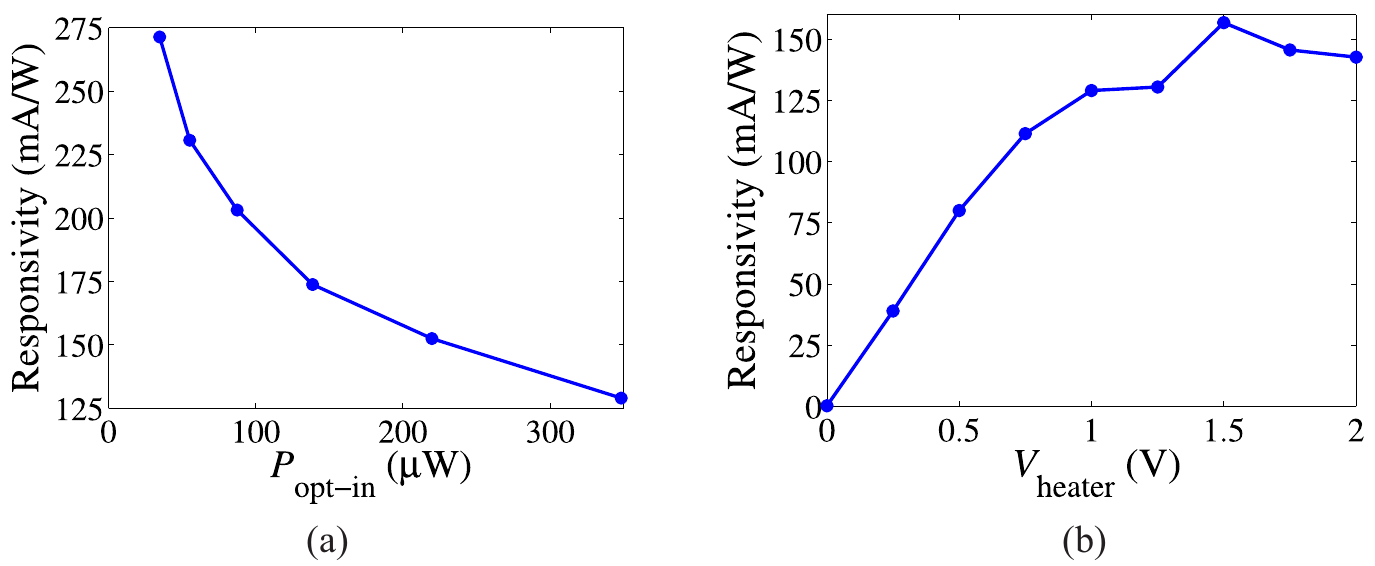}}
}
\caption{Responsivity of the IRPH, measured at the resonance wavelength of the MRR, (a) as a function of $P_{\text{opt-in}}$, with $V_{\text{heater}} = \SI{1}{V}$, and (b) as a function of $V_{\text{heater}}$, with $P_{\text{opt-in}} = \SI{348}{\uW}$.
}
\label{fig_resp}
\end{figure*}

The responsivity of the IRPH at the MRR's resonance wavelength, $\lambda_r$, is defined as $I_{\text{PD}} (\lambda_r)/P_{\text{opt-in}}$.
$P_{\text{opt-in}}$ is the optical power in the bus waveguide at the input of the MRR filter, which was estimated using the measured off-resonance through-port transmission. 
The measured responsivities of the IRPH as a function of $P_{\text{opt-in}}$ and $V_{\text{heater}}$ are shown in Figs.~\ref{fig_resp}(a) and \ref{fig_resp}(b), respectively.  
The measurements show that the responsivity decreases with increasing $P_{\text{opt-in}}$. 
We suspect that this is due to defect states being depleted faster than their replenishment by trapping carriers.
As $V_\text{heater}$ is increased [Fig.~\ref{fig_resp}(b)], the strength of the electric field is also increased. As a result, the carrier transit times across the waveguide is reduced in comparison to the carrier lifetimes. Consequently, the responsivity of the device is increased. However, at larger values of $V_\text{heater}$, velocity saturation of the carriers limits the overall responsivity. Velocity saturation has a similar effect on $I_\text{heater}$, as seen from Fig.~\ref{fig2}(a).
The high responsivities of our devices are partly due to the high optical intensity build-up inside the MRRs, which was calculated to be about $17.3 \times P_{\text{opt-in}}$. Since the intensity build-up factor of an MRR is proportional to the MRR's finesse parameter \cite{logan2011}, the MRRs used in this work were designed to have a maximized finesse as detailed in \cite{hasi2015}.
The typical responsivities measured for the IRPHs were about an order of magnitude larger than those reported for doped silicon waveguide PDs with $p\text{-}n$ \cite{haike2014, yu2012} or $p^{++}\text{-}i\text{-}p^{++}$ doping \cite{zhou2014}, and are of the same order as those reported in \cite{logan2011}, which required additional fabrication steps for defect implantation, and those reported in \cite{li2013}, which required high reverse bias voltages.

\section{Automated wavelength stabilization of an MRR filter}

\begin{figure*}[!h]
\centering
\centerline{
{\includegraphics[width=\textwidth]{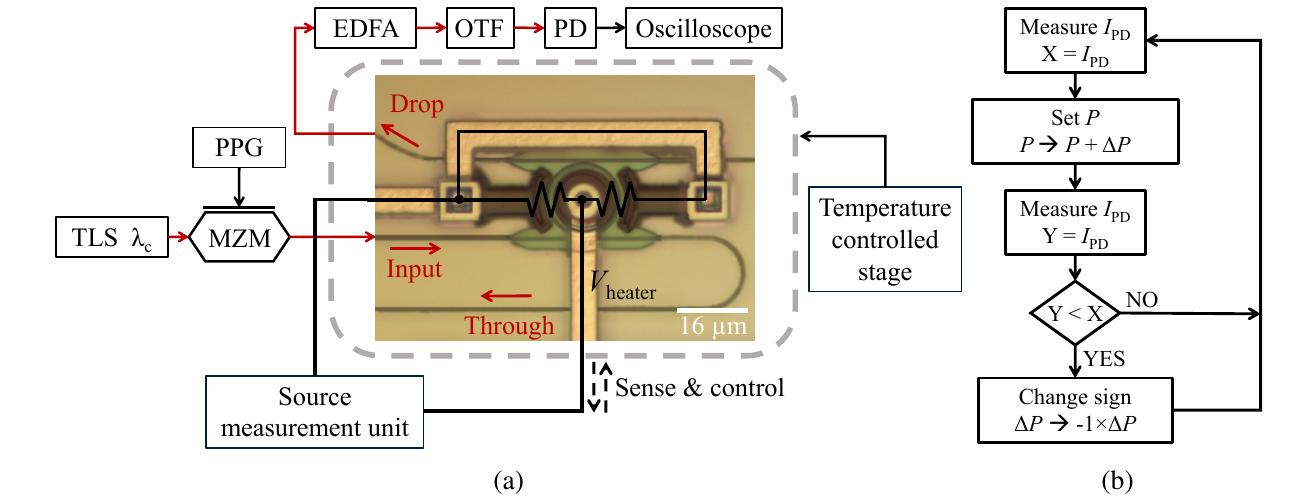}} 
}
\caption{(a) Schematic of the experimental setup. (b) Flow diagram of the control algorithm.
}
\label{mr_setup}
\end{figure*}

A schematic of the experimental setup used to demonstrate wavelength stabilization is shown in Fig. \ref{mr_setup}(a). The optical data stream was generated by modulating the output of a tunable laser source (TLS) using a LiNbO$_{3}$ Mach-Zehnder modulator (MZM). A pulse pattern generator (PPG) outputting a non-return-to-zero  2$^{31}$-1 pseudo random binary sequence at 12.5 Gb/s served as a modulating sequence. Grating couplers were used to couple light in to and out of the chip.  The output from the drop-port of the MRR filter was amplifed using an erbium-doped fiber amplifier (EDFA) and was filtered using an optical tunable filter (OTF). The output of the OTF was connected to a PD, which was connected to a real-time oscilloscope (\SI{32}{GHz}, \SI{80}{GSa/s}) for monitoring eye diagrams. The chip was mounted on a temperature controlled stage, which used a thermoelectric cooler (TEC) and thermistor combination to monitor and control the chip temperature. The sense and control operations on the IRPH were performed using a source measurement unit with a \SI{1}{\uA} current measurement resolution when used in constant voltage mode.

As shown in Fig. \ref{fig2}(b), the photocurrent generated in the IRPH is maximized when the MRR is on resonance. 
Therefore, wavelength stabilization was achieved using a computer implemented control algorithm similar to that in \cite{timurdogan2012}. The algorithm continuously searches for the $V_{\text{heater}}$ that maximizes $I_{\text{PD}}$. The flow diagram of the control algorithm is outlined Fig. \ref{mr_setup}(b), in which $P$ is the electrical power supplied to the heater. In each iteration of the algorithm, $V_{\text{heater}}$ is changed such that $P$ is changed by $\Delta P$. 
This ensures that the round-trip phase step, proportional to electrical power, does not depend on the operating voltage. $I_{\text{PD}}$ is measured before and after changing $P$. If $I_{\text{PD}}$ increases, then the sign of $\Delta P$ is unchanged and the algorithm proceeds to the next iteration. Otherwise, the sign of $\Delta P$ is reversed before proceeding.
A one-time calibration step is performed prior to starting the control algorithm to measure $I_{\text{heater}}$ as a function of $V_{\text{heater}}$ so that $I_{\text{PD}}$ can be calculated.

\begin{figure}[h]
\centering
\centerline{
{\includegraphics[width=\textwidth]{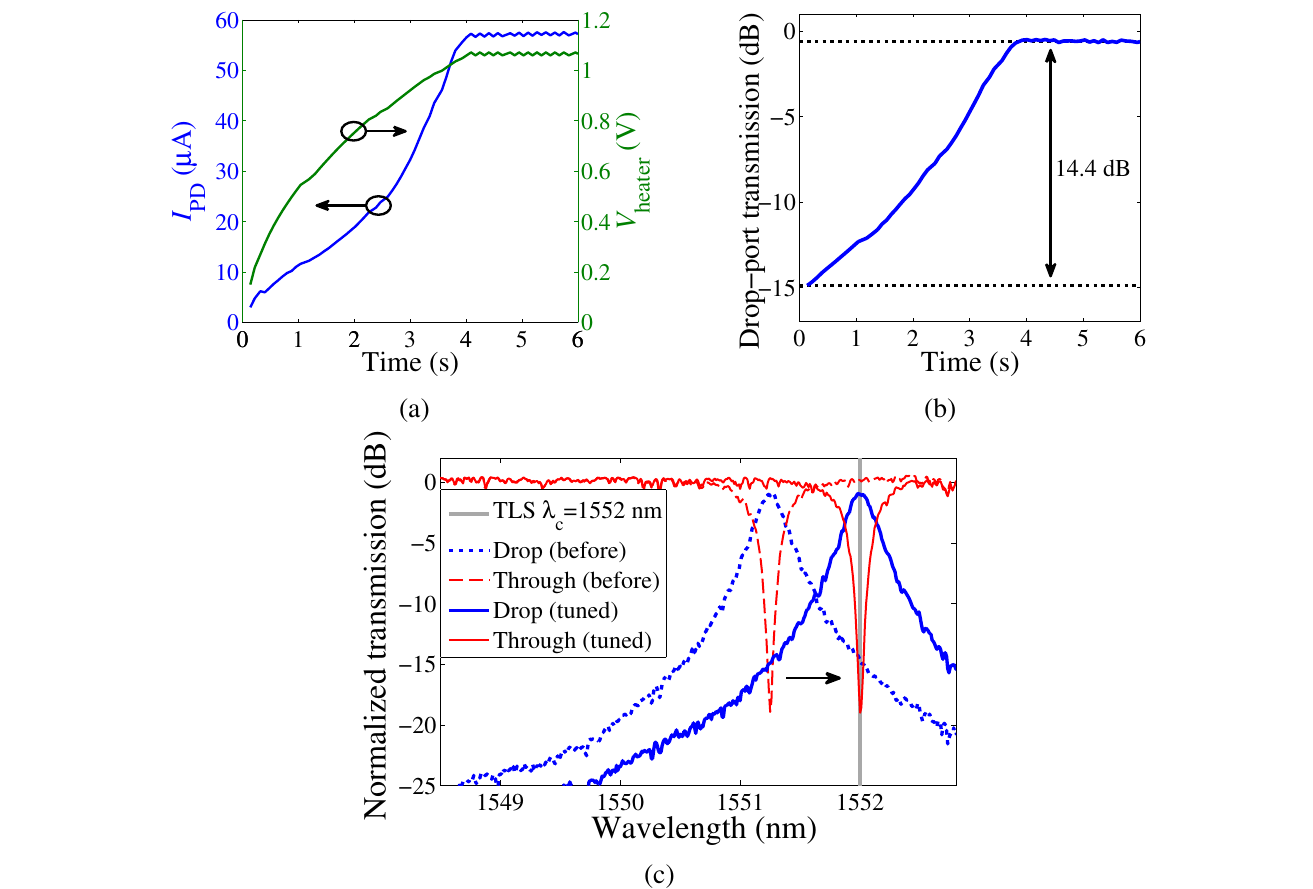}}
}
\caption{Measured (a) $I_{\text{PD}}$ and $V_{\text{heater}}$, and (b) normalized drop-port transmission, during the progression of the control algorithm. (c) Normalized  through- and drop-port transmission spectra of the MRR filter before and after the control algorithm was applied.
}
\label{mr_tuning}
\end{figure}

\begin{figure*}[!hb]
\centering
\centerline{
{\includegraphics[width=0.75\textwidth]{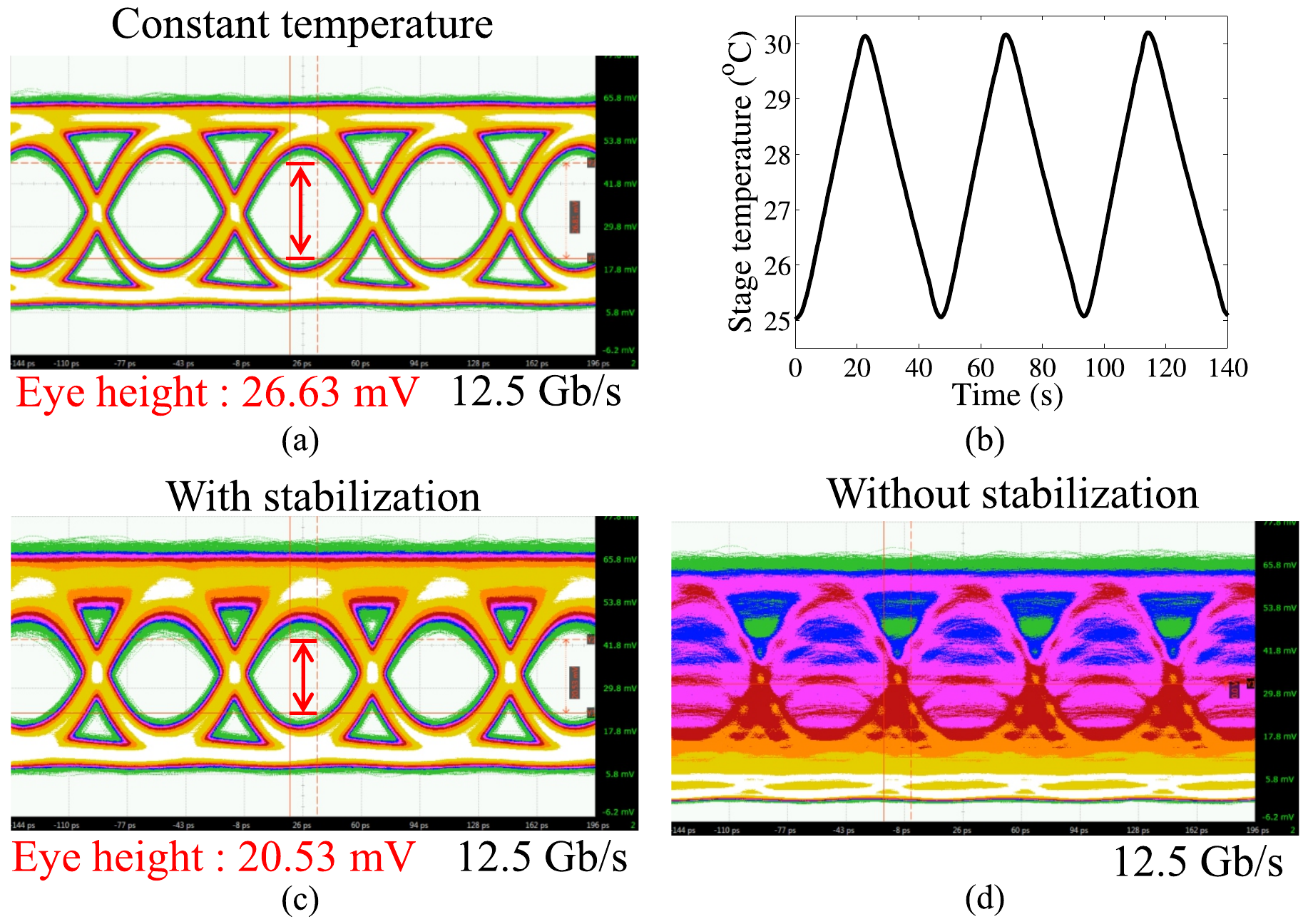}} 
}
\caption{(a) Measured eye diagram with a constant stage temperature. (b) Measured stage temperature as a function of time. Measured eye diagrams (c) with and (d) without automated wavelength stabilization. 
}
\label{mr_eye}
\end{figure*}

The photocurrent, heater voltage, and the measured drop-port optical power during the progression of the control algorithm are shown in Figs.~\ref{mr_tuning}(a) and \ref{mr_tuning}(b), respectively. 
The slow response time observed was primarily due to the high integration time of the off-chip optical power monitor that was used to measure the drop-port output power shown in Fig. \ref{mr_tuning}(b).
The recorded wavelength spectra in Fig. \ref{mr_tuning}(c) show that the drop-port power is maximized at the TLS output wavelength (\SI{1552}{nm}) after applying the control algorithm.

Eye diagram measurements were performed to demonstrate the quality of a transmitted signal through the MRR filter while the chip was subjected to temperature variations. Figure \ref{mr_eye}(a) shows the output eye diagram when the stage temperature was maintained at a constant level. For this measurement, the resonance wavelength of the MRR was manually aligned to the TLS wavelength and the control algorithm was not used.  
When the control algorithm was turned on and the stage temperature was kept constant, the change in eye height was negligible.
The applied stage temperature variation is shown in  Fig. \ref{mr_eye}(b). The temperature of the stage was changed from $\SI{25.0}{^\circ C}$ to $\SI{30.2}{^\circ C}$ and the rate of temperature change was approximately $0.28 ~ ^\circ$C/s. This corresponded to a resonance wavelength shift of approximately \SI{0.4}{nm} at a rate of about \SI{20}{pm/s}. 
The linewidth of the MRR filter was \SI{0.31}{nm}.
Figure \ref{mr_eye}(c) shows the eye diagram with wavelength stabilization while the stage temperature was varied as shown in Fig. \ref{mr_eye}(b). The eye remained open for the duration of the measurement. The minimum recorded eye height was about 23\% smaller than the eye height recorded at a constant temperature.
The eye height was reduced primarily because the speed of the optimization algorithm was limited by the response time of the measurement equipment.
As shown in Fig. \ref{mr_eye}(d), the eye diagram is completely closed when the wavelength stabilization was not used.
For each measurement, more than 1.7 $\times 10 ^{12}$ bits were transmitted.

\section{Automated wavelength tuning and stabilization of a second-order MRR filter} 

\begin{figure*}[!h]
\centering
\centerline{
{\includegraphics[width=0.95\textwidth]{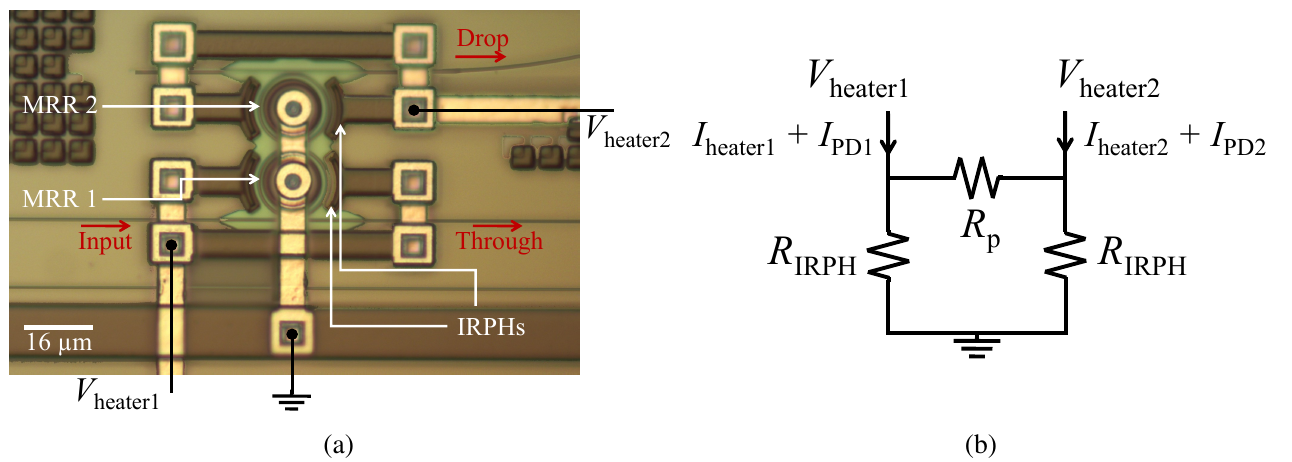}}
} 
\caption{(a) Picture of a fabricated second-order series-coupled MRR filter.
(b) Equivalent circuit diagram of the device.
}
\label{mr2_picture}
\end{figure*}

Compared to first-order filters, second-order series-coupled MRR filters offer superior spectral characteristics, such as a flat-top response and a high out-of-band signal rejection \cite{de_heyn2013, dahlem2011, boeck2011}.
A microscope image of such a filter is shown in Fig. \ref{mr2_picture}, where MRRs 1 and 2 are coupled to the through- and drop-port waveguides, respectively.
Ideally, for the desired filter response, both of the MRRs of a second-order filter should be resonant at the laser's wavelength, $\lambda_{\text{c}}$.  
However, fabrication and temperature variations cause the resonance wavelengths of the MRRs to be detuned from  each other, as well as from $\lambda_{\text{c}}$. 
Therefore, a control algorithm should be able to (1) automatically tune the MRRs to be resonant at $\lambda_{\text{c}}$, and (2) maintain the resonance at $\lambda_{\text{c}}$ while the chip is subjected to temperature variations.
In this section, we experimentally demonstrate the automated tuning of a second-order filter by sequentially tuning the MRRs to $\lambda_{\text{c}}$. The resonance conditions for the MRRs are found by maximizing $I_{\text{PD}1}$ and $I_{\text{PD}2}$, the photocurrents generated by the IRPHs integrated into MRRs 1 and 2, respectively.
We also demonstrate wavelength stabilization of the filter by using a control algorithm that monitors $I_{\text{PD}2}$ to maximize the drop-port output. 

The second-order filter was designed to have power coupling coefficients of $|\kappa|^2 =$~0.1311 and 0.0049, for the bus-to-MRR and MRR-to-MRR couplers, respectively.
The other parameters used for the design of the MRRs and the IRPHs were similar to those used for the first-order MRR design described in section 2. MRRs 1 and 2 can be tuned by adjusting $V_{\text{heater1}}$ and $V_{\text{heater2}}$, respectively. 
An equivalent electrical circuit of the device is shown in Fig. \ref{mr2_picture}(b). 
$R_{\text{p}}$ is a parasitic resistance due to the silicon slab region between MRRs 1 and 2, and was measured to be about \SI{3}{k$\Omega$}, which was approximately $10 \times $ the resistance of the IRPHs,  $R_{\text{IRPH}}$. 
Due to the presence of  $R_\text{p}$, $I_{\text{heater1}}$ and $I_{\text{heater2}}$ each depended on the voltages applied to both of the IRPHs.
Therefore, to calculate $I_{\text{PD1}}$ and $I_{\text{PD2}}$ from the total measured current, we  performed a 2-D calibration measurement of 
$I_{\text{heater1}}$ and $I_{\text{heater2}}$ as functions of both $V_{\text{heater1}}$ and $V_{\text{heater2}}$. However, the 2-D calibration could be avoided by (1) characterizing $R_{\text{IRPH}}$ and $R_{\text{p}}$ as functions of the voltages across them, and using the results to calculate  $I_{\text{heater1}}$ and $I_{\text{heater2}}$ as functions of $V_{\text{heater1}}$ and $V_{\text{heater2}}$ or (2) designing the device such that the coupling between MRRs 1 and 2 is achieved without a slab region, i.e. with strip waveguides.

\begin{figure*}[t]
\centering
\centerline{
{\includegraphics[width=\textwidth]{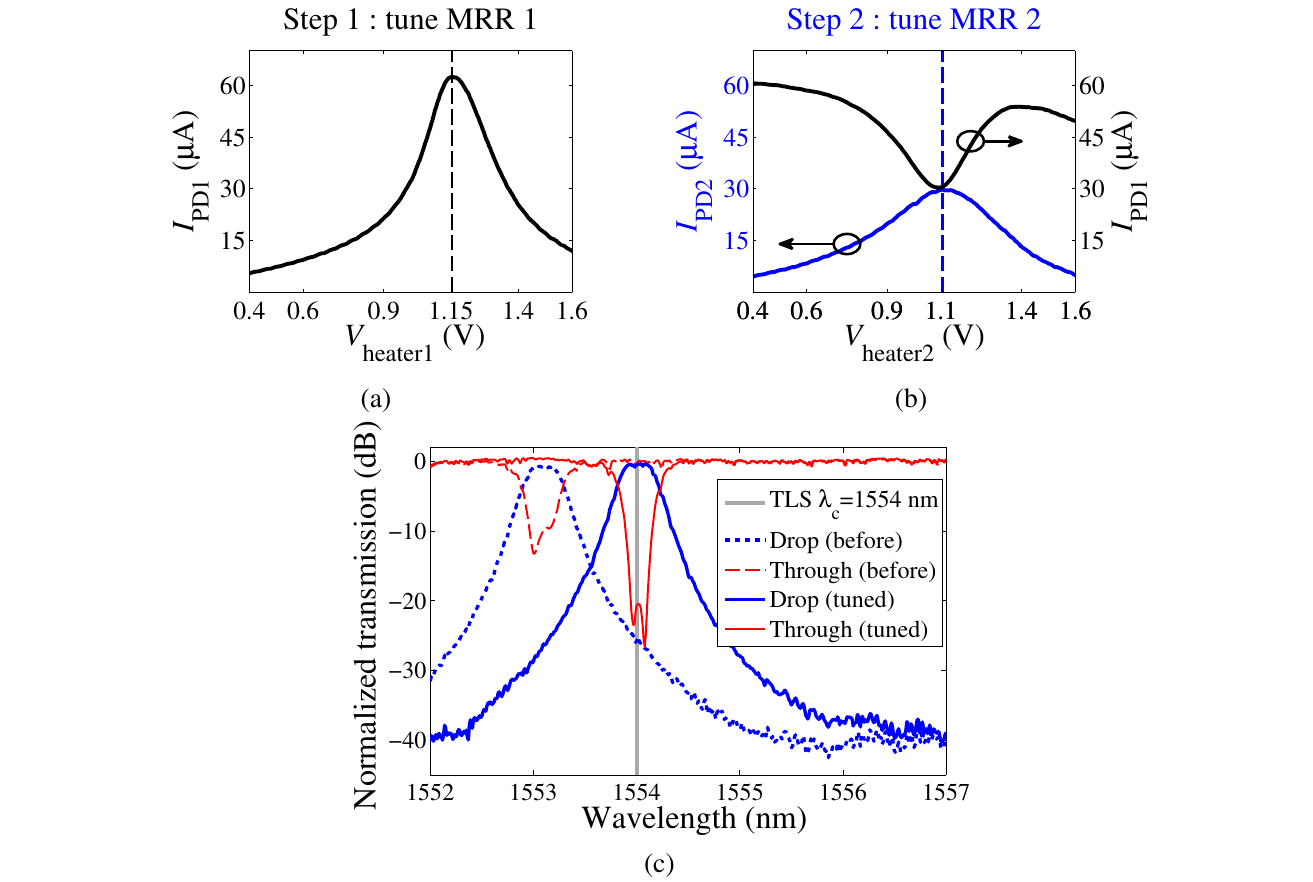}} 
}
\caption{Measured (a) $I_{\text{PD1}}$ as a function of $V_\text{heater1}$, with $V_\text{heater2} =$ \SI{0}{V}, and (b) $I_{\text{PD1}}$ and $I_{\text{PD2}}$ as functions of $V_\text{heater2}$, with $V_\text{heater1} =$ \SI{1.15}{V}. (c) Measured through- and drop-port responses of the second-order series-coupled MRR filter before and after tuning.
}
\label{mr2_tuning}
\end{figure*}

The tuning process is shown in Fig. \ref{mr2_tuning}. 
First,  as shown in Fig. \ref{mr2_tuning}(a), $I_{\text{PD1}}$ was measured as a function of $V_{\text{heater1}}$ with $V_{\text{heater2}}=0$.
To tune MRR 1 to $\lambda_{\text{c}}$,  $V_{\text{heater1}}$ was set to \SI{1.15}{V}, the value that maximized $I_{\text{PD1}}$. 
Next, as shown in Fig. \ref{mr2_tuning}(b), $I_{\text{PD2}}$ was measured as a function of $V_{\text{heater2}}$ with $V_{\text{heater1}} = $ \SI{1.15}{V}.
To tune MRR 2 to $\lambda_{\text{c}}$,  $V_{\text{heater2}}$ was set to \SI{1.10}{V}, the value that maximized $I_{\text{PD2}}$. 
Figure \ref{mr2_tuning}(b) also shows the change in $I_{\text{PD1}}$, which corresponds to the change in light intensity in MRR 1, while tuning MRR 2.
The through- and drop-port spectra before and after tuning the filter are shown in Fig. \ref{mr2_tuning}(c). After tuning, the spectral response of the filter was improved, with a drop-port insertion loss less than 0.5 dB and a through-port extinction ratio greater than \SI{20}{dB}.

\begin{figure*}[h]
\centering
\centerline{
{\includegraphics[height=2in]{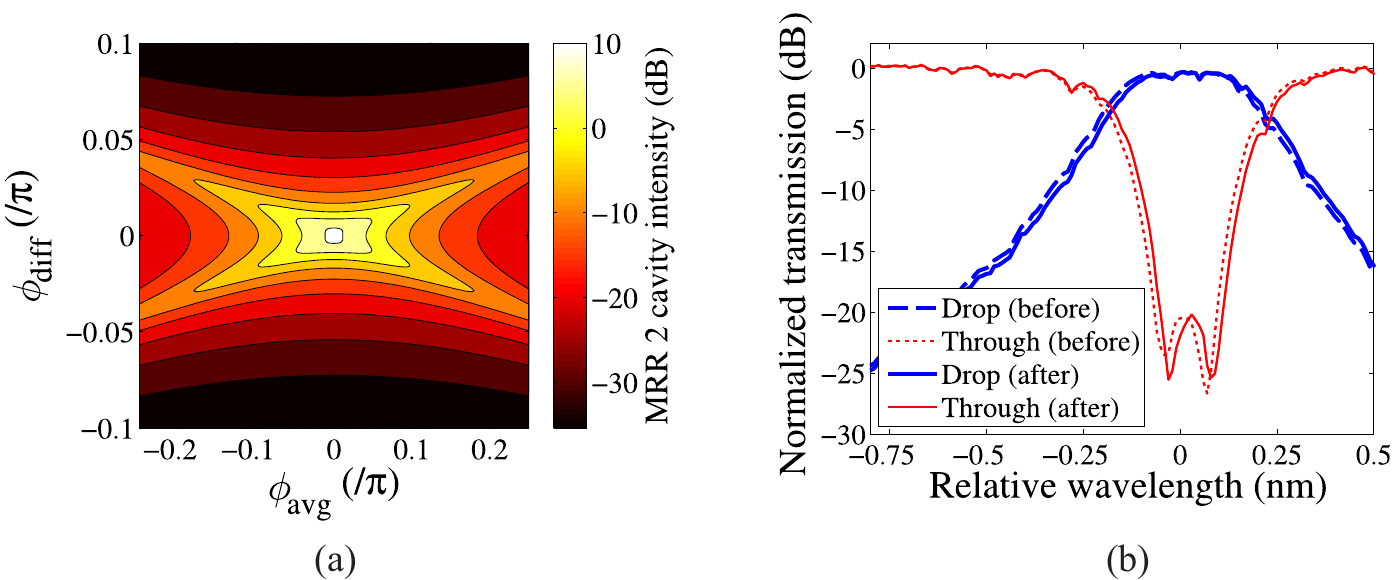}}
}
\caption{(a) Calculated normalized cavity intensity in MRR 2 as function of $\phi_{\text{diff}}$ and $\phi_{\text{avg}}$. (b) Through- and drop-port transmission spectra before and after applying the control algorithm to the initially tuned spectrum in Fig. \ref{mr2_tuning}(c). Wavelength is relative to \SI{1554}{nm}.
}
\label{mr2_algorithm}
\end{figure*}

\begin{figure*}[h]
\centering
\centerline{
{\includegraphics[width=0.9\textwidth]{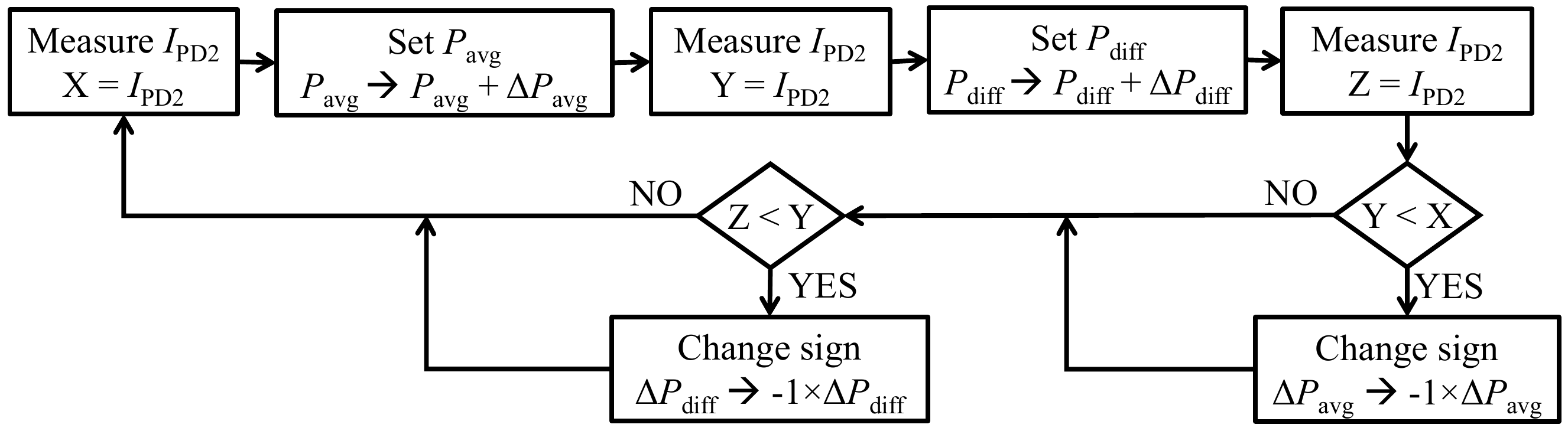}}
}
\caption{Flow diagram illustrating the the control algorithm used to stabilize the second-order series-coupled MRR.
}
\label{mr2_flow}
\end{figure*}

The algorithm used for wavelength stabilization can be explained by expressing the tuning states of the MRRs in terms of the transformed coordinates 
${\phi_{\text{diff}} = (\phi_1 - \phi_2)/2}$ and ${\phi_{\text{avg}} = (\phi_1 + \phi_2)/2}$, where $\phi_1$ and $\phi_2$ are the round-trip phases of MRRs 1 and 2, respectively.
The round-trip phases can be tuned by changing $P_1$ and $P_2$, the electrical powers supplied to the IRPHs in MRRs 1 and 2, respectively.
Figure \ref{mr2_algorithm}(a) shows the optical cavity intensity in MRR 2 as a function of $\phi_{\text{diff}}$ and $\phi_{\text{avg}}$, calculated using the transfer matrix method described in \cite{poon2004}. 
The intensity, being proportional to the filter's drop-port output intensity, has a single maximum point corresponding to the case $\phi_1 = \phi_2 = 0$, i.e. ${\phi_{\text{diff}} = \phi_{\text{avg}} = 0}$, which yields the tuned filter response at $\lambda_{\text{c}}$. 
Therefore, a control algorithm for wavelength stabilization using IRPHs can be based on the maximization of $I_{\text{PD2}}$, thereby maximizing the light intensity in MRR 2.
A flow diagram of the implemented control algorithm is shown in Fig. \ref{mr2_flow}.
The control algorithm continuously updated (1) the average power supplied to the heaters, ${P_{\text{avg}} = (P_1 + P_2)/2}$, by a step size of ${\pm \Delta P_{\text{avg}}}$ and (2) the difference in power supplied to the heaters, ${P_{\text{diff}} = (P_1 - P_2)/2}$, by a step size of 
${\pm \Delta P_{\text{diff}}}$. The signs of $\Delta P_{\text{avg}}$ and $\Delta P_{\text{diff}}$ were chosen in each iteration of the algorithm towards increasing $I_{\text{PD2}}$. 
The MRRs were tuned in terms of $P_{\text{diff}}$ and $P_{\text{avg}}$ because the largest slopes of MRR 2's optical cavity intensity near its maximum occur along the $\phi_{\text{diff}}$ and $\phi_{\text{avg}}$ axes [Fig. \ref{mr2_algorithm}(a)], thereby yielding the highest sensitivity to detuning when a maximum search algorithm is used. Alternatively, one could implement the control algorithm by tuning the MRRs in terms of $P_1$ and $P_2$.
Figure \ref{mr2_algorithm}(b) shows the measured spectra before and after applying the control algorithm to the filter while keeping the temperature of the stage constant. Before applying the control algorithm, the filter was tuned to a wavelength of \SI{1554}{nm} using the wavelength tuning algorithm. It can be seen that the optimized state of the filter as determined by the tuning algorithm agrees with that determined by the stabilization algorithm.

\begin{figure*}[!t]
\centering
\centerline{
{\includegraphics[width=0.75\textwidth]{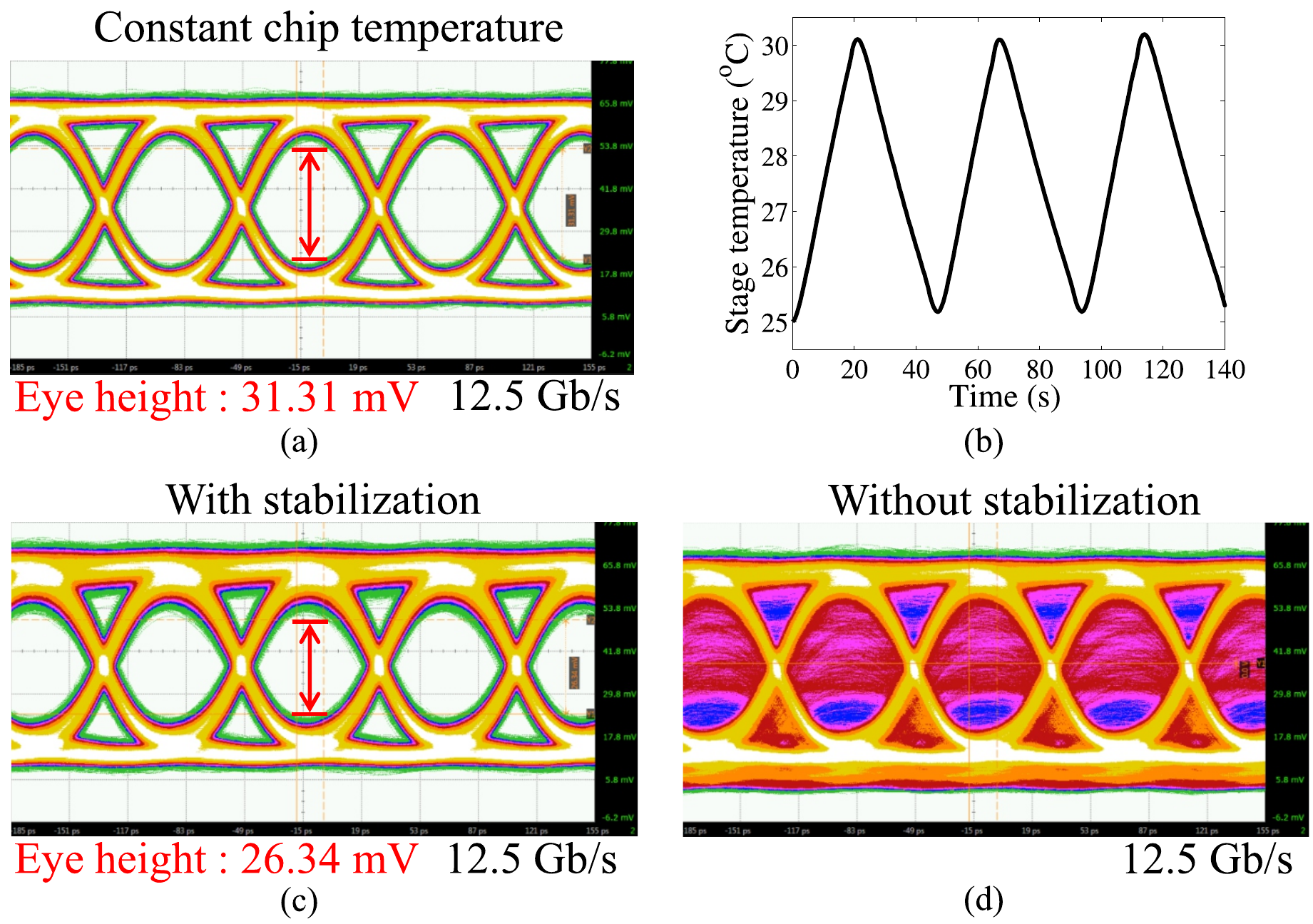}} 
}
\caption{(a) Measured eye diagram at constant stage temperature. (b) Measured stage temperature as a function of time. Measured eye diagrams (c) with and (d) without automated wavelength stabilization. 
}
\label{mr2_eye}
\end{figure*}

We used a similar setup to the one shown in Fig. \ref{mr_setup} to perform eye diagram measurements.
Figure \ref{mr2_eye}(a) shows the output eye diagram when the stage temperature was maintained at a constant level and the wavelength stabilization was not used. 
When the wavelength stabilization was used and the stage temperature was kept constant, the eye height decreased by less than 3\%. The reduction in eye height was due to oscillations about the optimal tuning point caused by the electrical power steps in each iteration of the stabilization algorithm.
The stage temperature was varied according to the profile shown in Fig. \ref{mr2_eye}(b) in order to test the wavelength stabilization algorithm.
The temperature of the stage was changed from $\SI{25.0}{^\circ C}$ to $\SI{30.2}{^\circ C}$ and the rate of temperature change was approximately 0.28 $^\circ$C/s.
The linewidth of the filter was \SI{0.39}{nm}.
The eye which remained open with wavelength stabilization, as shown in  Fig. \ref{mr2_eye}(c), was completely closed  when wavelength stabilization was not used [Fig. \ref{mr2_eye}(d)]. 
For each measurement, more than 1.7 $\times 10 ^{12}$ bits were transmitted.
With wavelength stabilization, the minimum recorded eye height was about 16\% smaller than the eye height recorded at a constant temperature. This reduction in eye height was primarily due to the time required to optimize the tuning state after a temperature change resulted in a detuning.

\section{Automated wavelength tuning of higher-order series-coupled MRR filters}

\begin{figure*}[!b]
\centering
\centerline{
{\includegraphics[width=0.98\textwidth]{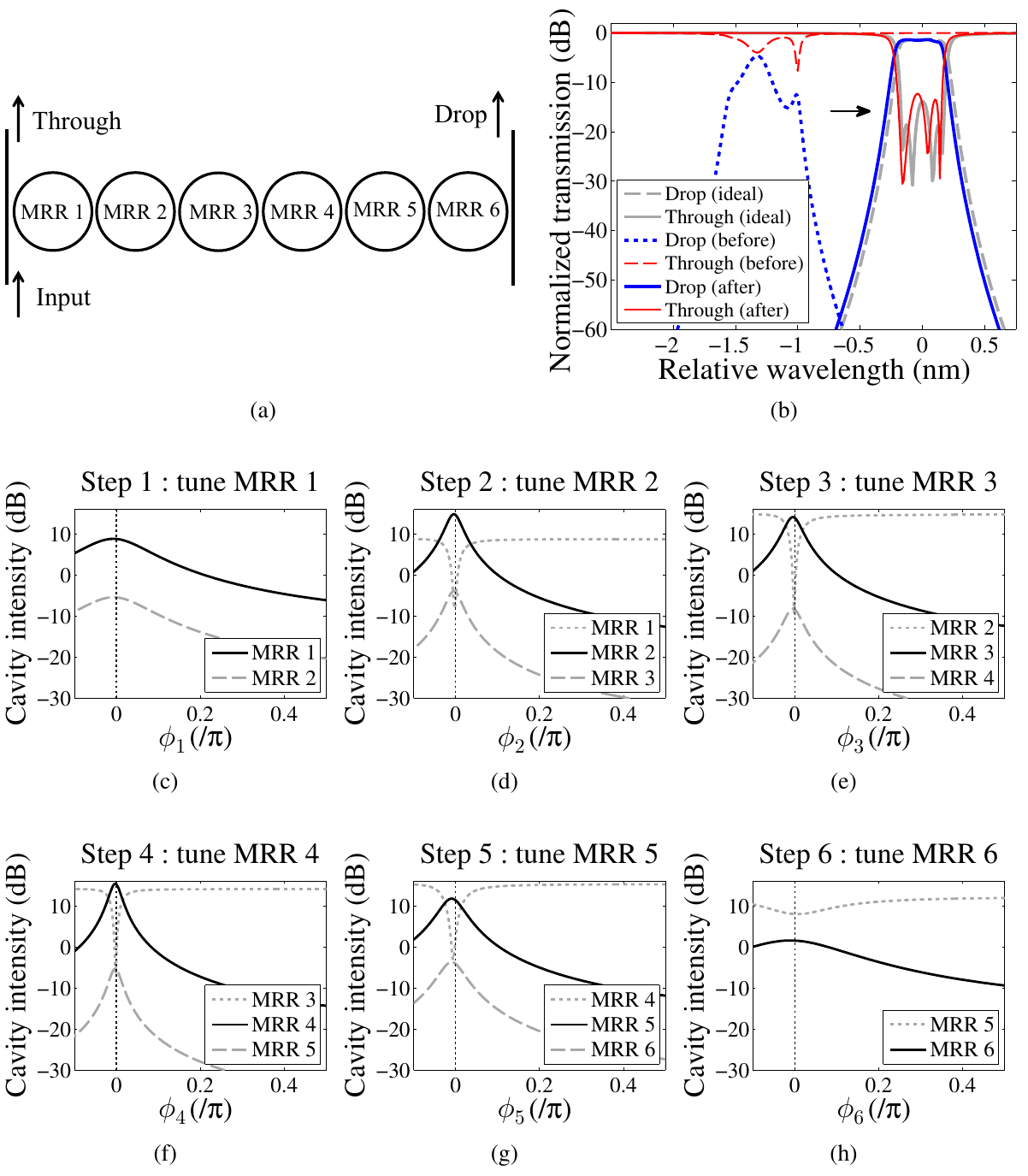}}
}
\caption{(a) Schematic of a series-coupled 6$^{\text{th}}$-order MRR filter.  (b) Simulated through- and drop-port transmission spectra  before and after  tuning. (c)-(h) Calculated  optical cavity intensity in MRRs 1-6 as a function of $\phi_{1-6}$ as the MRRs are sequentially tuned to $\lambda_{\text{c}}$. The optical transmission in (b) and the cavity intensities in (c)-(h) are all normalized to the optical intensity at the input of the filter.
}
\label{fig_high}
\end{figure*}

Tuning higher-order MRR filters by only monitoring the drop-port intensity can be challenging because information about the round-trip phases of each MRR can not be determined independently from the drop-port transmission alone. Therefore, tuning algorithms for higher-order series-coupled MRR filters based on drop-port intensity monitoring have relied on complex optimization algorithms \cite{mak2015}. 
In this section, we theoretically show that the automated tuning method based on monitoring the intensity in each MRR using IRPHs, described in section 4 for second-order filters, can be readily extended to series-coupled MRR filters with a higher number of MRRs.

As an example, we consider the 6$^{\text{th}}$-order series-coupled add/drop MRR filter shown in Fig. \ref{fig_high}(a).  For a maximally flat pass-band response \cite{little1997},  the power cross-coupling coefficients for each of the couplers from the input- to the drop-port were chosen to be ${|\kappa|^2 = 0.4, 0.0146, 0.0039, 0.0029, 0.0039, 0.0146, \text{ and } 0.4,}$ respectively. 
A radius of $\SI{8}{\um}$, effective index of $2.57$, and a waveguide loss of \SI{6}{dB/cm} were assumed for all of the MRRs. 

Figure \ref{fig_high}(b) shows the calculated filter responses before and after tuning. The wavelength is shown relative to $\lambda_{\text{c}} = \SI{1554.2}{nm}$. 
In order to simulate the effects of fabrication variations, the round-trip phase for the $n^\text{th}$ MRR in the filter, $\phi_n$, was calculated using ${\phi_n = \phi_{{0,}n}+\phi_{\text{th},n}}$,
where $\phi_{0,n}$ is the initial round-trip phase at $\lambda_{\text{c}}$ and $\phi_{\text{th},n}$ is the phase associated with the thermal tuning. $\phi_{0,n}$ was modeled using a normal distribution with a mean of $-\pi/5$ and a standard deviation of 0.0713$\pi$. The mean phase of $-\pi/5$ was chosen to ensure that all of the rings are initially detuned from $\lambda_{\text{c}}$. The value for standard deviation was calculated according to the fabrication variations in MRR resonance wavelengths as reported in \cite{lukas2014}. The tuning algorithm described herein is not sensitive to the values of these parameters as long as the rings are all initially sufficiently detuned from $\lambda_{\text{c}}$.
The ideal filter response, shown in Fig. \ref{fig_high}(b), corresponds to the case in which each $\phi_{n} = 0$. 
Therefore, the goal of the tuning method is to adjust each $\phi_{\text{th},n}$ such that the ideal filter response is achieved.

Automatic tuning of the filter can be achieved by sequentially tuning MRRs 1 through 6 to be resonant at $\lambda_{\text{c}}$. 
Figure \ref{fig_high}(c) shows the optical intensity in MRRs 1 and 2 as $\phi_1$ is tuned. 
The value of $\phi_1$ corresponding to the maximum optical intensity in MRR 1, $\phi_{\text{1,max}}$, is determined and $\phi_1$ is set to this value. 
Then, similarly, MRRs 2 through 6 are sequentially tuned, and each $\phi_n$ is set to $\phi_{n,\text{max}}$.
Figures~\ref{fig_high}(c)-\ref{fig_high}(h) show, for each step of the tuning process, the optical cavity intensity in the MRR being tuned as well as the intensities in the adjacent MRRs. 
As shown in Fig. \ref{fig_high}(b), the filter spectra after tuning show excellent agreement with the ideal response. 
As it was experimentally demonstrated in section 4 for a second-order filter, the phases $\phi_n$ are tuned by changing the electrical power supplied to the IRPH in each MRR, and the intensity in each ring can be determined by measuring $I_{\text{PD},n}$. 
In Figs.~\ref{fig_high}(c)-\ref{fig_high}(h), the maximum optical cavity intensity in the $n^{\text{th}}$ MRR during tuning step $n$ does not occur exactly at ${\phi_n = 0}$ due to the influence of the resonances of the other MRRs. The accuracy of the algorithm is best when the initial detuning of the resonators is increased. Here, a mean initial detuning of $-\pi/5$ was chosen to demonstrate that excellent agreement with the ideal filter spectrum can be obtained, even for a small initial detuning.
This method of tuning can be generalized to series-coupled filters with any number of MRRs, and the time required for tuning would scale linearly with the order of the filter.  
For the stabilization of a high-order MRR-based filter, one can extend the control algorithm we used for the second-order ring (Fig.~\ref{mr2_flow}). The algorithm would consist of sequentially stepping the electrical power supplied to each of the IRPHs in the direction that maximizes the photocurrent in the IRPH of the MRR coupled to the drop-port waveguide.

\section{Discussion and conclusions}

IRPHs use doped waveguides, which have additional losses compared to undoped waveguides. Nevertheless, the MRR devices can still be designed to have low drop-port losses by controlling the bus-to-MRR and MRR-to-MRR coupling coefficients. 
PDs based on $p\text{-}n$ junctions \cite{li2015, yu2012} operate in reverse bias and therefore have low dark currents. The IRPHs demonstrated here have dark currents (i.e. $I_\text{heater}$) on the order of milliamps. 
However, since this current is simultaneously used for the TO tuning of the MRR, the dark current does not represent an additional power consumption. 
The measurements showed that the responsivity of an IRPH is a function of the input power [Fig.~\ref{fig_resp}(a)] and of the bias voltage [Fig.~\ref{fig_resp}(b)]. 
However, it was found that these effects did not have an appreciable impact on the performance of wavelength tuning and stabilization.
The responsivities of the IRPHs are large enough so that they can be used for wavelength tuning and stabilization even at low voltages.  We observed successful wavelength stabilization for voltages as low as \SI{0.3}{V}.
The calibration step required for measuring $I_{\text{PD}}$ was only performed at the start of the measurement, which was then used for demonstrating wavelength stabilization for a temperature range of 5~$^\circ$C.
Beyond this range, we were unable to maintain the rate of change of the temperature due to limitations in our experimental setup. It should be noted that the initial calibration may not remain valid for large temperature variations as the resistance of the IRPH is temperature dependent, and could require recalibration. 
The 2-D calibration we performed for the second-order MRR does not scale well for higher-order filters. However, the calibration complexity can be made simple, i.e. to scale linearly with filter order, by eliminating the parasitic resistances between the MRRs. 
As suggested in section 4, this could be achieved by implementing MRR-to-MRR couplers with strip waveguides, thereby avoiding the conductive silicon slab region between the MRRs.  
In our experiments, the feedback loops were implemented using a source measurement unit and a computer. The response times of the feedback loops were limited due to the slow response of this system, and not due to the response time of the IRPHs. It has been previously reported that photodetection based on DSA can be fast, allowing for signal detection at multi-Gb/s datarates \cite{yu2012}.

In summary, we have demonstrated automated wavelength tuning and stabilization of MRR-based filters using IRPHs. 
The responsivities measured for the IRPHs were as high as \SI{271}{mA/W}, where $V_\text{heater} = $ \SI{1}{V} and the intensity build-up factor of the MRRs was about 17.3.
The IRPHs measure the light intensity in the MRRs and, therefore, can be used for automated tuning of MRRs. 
We implemented wavelength tuning and stabilization algorithms for first-order and second-order series-coupled MRR filters, and obtained open eye diagrams for both cases at a datarate of \SI{12.5}{Gb/s} while varying the temperature. Also, we theoretically showed how higher-order series-coupled MRR filters can be automatically tuned using IRPHs by sequentially aligning the resonance wavelength of each MRR to the laser's wavelength. 
As demonstrated in this paper, the main advantages of using IRPHs for automatic wavelength tuning and  stabilization  are that   (1) they neither require any dedicated ion implantation steps to introduce defects, nor any germanium deposition \cite{li2015, poulton2015}, (2) they can be used for both photodetection and heating, which allows for a smaller footprint, and (3) they do not require tap-outs of the output signal \cite{padmaraju2014, cox2014, timurdogan2012, mak2015}.
Furthermore, since the IRPHs monitor the optical cavity intensity in the MRRs, this approach can be readily extended to devices/systems (e.g. \cite{ophir2013,chengli2014,dahlem2011}) which require simultaneous wavelength  tuning or stabilization of multiple MRRs.

\section*{Acknowledgments}
The authors would like to thank the Natural Sciences and Engineering Research Council of Canada, the Si-EPIC program, and CMC Microsystems for financial support. The authors also thank CMC Microsystems for providing design tools and support. The authors thank Dr. Roberto Rosales for helping with test equipment and Robert Boeck for helpful discussions.

\end{document}